# Imaging with Diffractive Axicons Rapidly Milled on Sapphire by Femtosecond Laser Ablation

Daniel Smith,[1,4] Soon Hock Ng,[1,4] Molong Han,[1,4] Tomas Katkus,[1] Vijayakumar Anand,[1,*] Karl Glazebrook,[2] and Saulius Juodkazis[1,3,]

[1]*Optical Sciences Centre and ARC Training Centre in Surface Engineering for Advanced Materials (SEAM), School of Science, Swinburne University of Technology, Hawthorn, VIC 3122, Australia.*
[2]*Centre for Astrophysics and Supercomputing, School of Science, Swinburne University of Technology, Hawthorn, VIC 3122, Australia.*
[3]*Tokyo Tech World Research Hub Initiative (WRHI), School of Materials and Chemical Technology, Tokyo Institute of Technology, 2-12-1, Ookayama, Meguro-ku, Tokyo 152-8550, Japan.*
[4]*The authors contributed equally to this work*
*\*vanand@swin.edu.au*

**Abstract:** Fabrication of large area (sub-1 cm cross-section) micro-optical components in a short period of time (~10 min) and with lesser number of processing steps is highly desirable and cost-effective. In the recent years, femtosecond laser fabrication technology has revolutionized the field of manufacturing by offering the above capabilities. In this study, a fundamental diffractive optical element, binary axicon – axicon with two phase or amplitude levels, has been designed in three configurations namely conventional axicon, photon sieve axicon (PSA) and sparse PSA and directly milled onto the Sapphire substrate. The fabrication results revealed that a single pulse burst fabrication can produce a flat and smooth profile than pulse overlapped fabrication which gives rise to surface damage and increased roughness. The fabricated elements were processed in IsoPropyl Alcohol and Potassium Hydroxide to remove debris and redeposited amorphous Sapphire. An incoherent illumination was used for optical testing of the components and a non-linear optical filter was used for cleaning the noisy images generated by the diffractive optical elements.

## 1. Introduction

Generation and precise control of optical fields is crucial in many optical instruments and imaging systems [1]. The optical fields can be engineered using optical components belonging to different categories such as refractive [2], diffractive [2], reflective [3], metaoptics [4-6] polarization optical components such as Q-plates [7] and liquid crystal optical components [8] based on the principle of operation. Depending upon the degree of control needed on the components of the optical field such as amplitude, phase and polarization, the complexity involved in design as well as fabrication varies. Next to the fully matured refractive optical technology, diffractive optics remains as the most widely used mode of beam control offering easier design, reasonable fabrication costs, and reaching diffraction efficiency closer to the values of refractive elements. Diffractive optics plays an important role in many areas of research such as optical trapping and tomography [9], computer generated holography [10], biomedical applications [11], integrated optics [12], display technology and serve also as components in augmented and mixed reality glasses [8].

The diffractive optical elements (DOEs) can be manufactured using different techniques such as photolithography [13], electron beam lithography [14], ion beam lithography [15], depending upon the feature sizes and area of the design. All the above methods are not only time-consuming but also have higher operating costs resulting in an increase in the cost of the DOEs. Except ion beam lithography, the other two methods cannot fabricate elements directly on to the substrate and therefore requires additional processing steps such as reactive ion etching to transfer the pattern from the resist layer to the substrate which increases the cost of manufacturing further. The electron beam and ion beam lithography methods are not suitable for fabrication of large area DOEs. Even though it may be argued that photolithography can transfer large area patterns, the time is still spent on fabrication of the mask. Therefore, except for mass production, even photolithography cannot be considered as a rapid fabrication system.

In the recent years, there has been a revolution in manufacturing of DOEs due to the advent of femtosecond ablation techniques [16-19]. They not only offer capabilities to manufacture large area DOEs but also can reach subwavelength features by the generation of harmonics. Furthermore, the vacuum conditions and other environmental isolation needed for electron and ion beam optics can be avoided in the femtosecond fabrication system. All the above makes femtosecond fabrication systems superior to the existing methods on various technological and economical fronts.

In this study, the femtosecond fabrication by ablation has been implemented to rapidly fabricate two-level axicons directly on to Sapphire substrates. The axicons were realized in conventional ring as well as sieve configurations. This manuscript consists of four sections. The calculation of diffracted intensity distribution and the simulation results are presented in the next section. The third section contains the fabrication procedure and characterization and optical experiments and beam cleaning procedure. The summary and conclusion are presented in the final section.

## 2. Methodology

In most of the earlier studies on beam characteristics of axicon only a coherent illumination is considered [20-22]. In this study, the beam characteristics are investigated for a spatially incoherent illumination as it is easily available at a low cost and will be highly relevant to large scale and astronomical applications where large area devices are often required which in turn can be manufactured using femtosecond fabrication systems. For a single point object, the behaviour of a coherent light is like that of an incoherent one [23]. However, when there is more than one point, then the optical system differs hugely between coherent and incoherent illuminations. The optical configuration is shown in Fig. 1 in which the light diffracted from a point object is incident on a the diffractive axicon and the intensity distribution is recorded. The distance between the object and the axicon is $u$ and the distance between the axicon and the sensor is $v$.

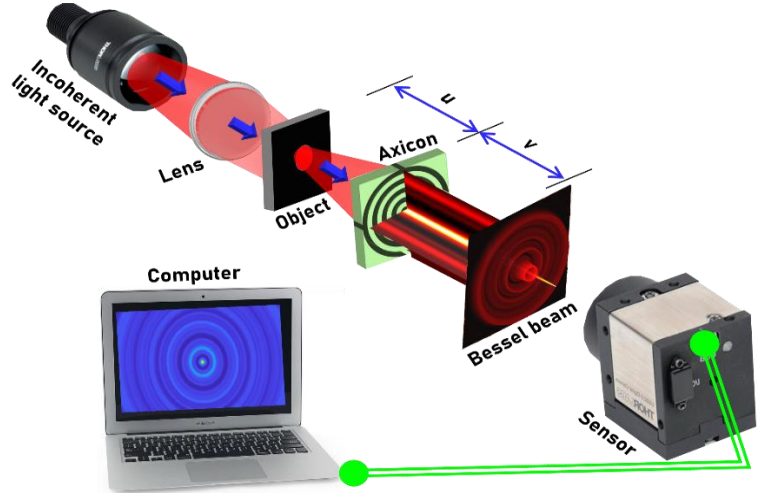

Fig. 1 Optical configuration for the generation of Bessel beams by a diffractive axicon

The binary axicon can be realized either as concentric rings or rings filled with discs similar to a photon sieve [24]. While the former with concentric rings has been widely used, only a few reports are available on axicons with sieve configuration. The rationale for comparing a conventional axicon with a photon sieve axicon (PSA) is that the fabrication using femtosecond ablation occurs point by point and it is much easier to realize an axicon as a PSA (disconnected discs) rather than in conventional type (overlapped discs). The radii of zones of a binary axicon can be given as

$$r_{1-n} = n\Lambda, \qquad (1)$$

where $\Lambda$ is the period of the axicon and $n$ is an integer. The phase of a binary axicon can be expressed as

$$\Phi_{Axicon} = \begin{cases} \pi, & r_{1-n+1/2} \leq R \leq r_{1-n+1} \\ 0, & elsewhere \end{cases}, \text{ where } R = \sqrt{x^2 + y^2}. \qquad (2)$$

The phase of a PSA is mathematically complicated as it is made up of circular discs of the same diameter instead of rings of varying diameters as in an axicon. A two-step process is proposed to design the PSA with polar coordinates $(R, \varphi)$ where $\varphi = tan^{-1}(y/x)$. In the first step, a mask consisting of Delta functions located at the $x$ and $y$ coordinates of the centers of the discs of diameter $\Lambda/2$ is generated. The centers of the circular discs can be located only at specific radial values given as

$$r_{2-n} = (4n-1)\Lambda/4. \tag{3}$$

The circular discs cannot be tightly arranged in the ring of an axicon as the circumference given as $2\pi r_{2-n}$ cannot be an integral multiple of $\Lambda$ for all values of $r_{2-n}$. The difference between the above two values is given as

$$L(r_{2-n}) = 2\pi r_{2-n} - \{floor(4\pi r_{2-n}/\Lambda)\}\Lambda/2, \tag{4}$$

and so, the increment in location needed for every disc is

$$\Delta(r_{2-n}) = L(r_{2-n})/\{floor(4\pi r_{2-n}/\Lambda)\}. \tag{5}$$

The values of the angles are given as

$$\theta_m(r_{2-n}) = [m\{\Lambda + \Delta(r_{2-n})\}]/r_{2-n}, \tag{6}$$

where $m$ varies from 0 to $floor(4\pi r_{2-n}/\Lambda)$. The mask consisting of Delta functions is given as

$$Mask = \begin{cases} \delta, R = r_{2-n} \text{ and } \varphi = \theta_m \\ 0 \qquad\qquad elsewhere \end{cases}. \tag{7}$$

At the end of the first step, the map of Delta functions is obtained for mounting the circular discs in the next step. To generate the mask of PSA ($M_{PSA}$), it is just sufficient to carry out a 2D convolution of the Mask function with a Circ function which is a circular top hat with a radius given by

$$M_{PSA} = Mask \otimes circ(\Lambda/2), \tag{8}$$

where '$\otimes$' is a 2D convolutional operator. The phase of the PSA is given as $\Phi_{PSA} = exp(j\pi M_{PSA})$.

Let us consider a point object located at a distance of $u$ from the diffractive axicon emitting light with an amplitude of $\sqrt{I_o}$. The complex amplitude reaching the axicon is given as

$$\psi_1 = C_1\sqrt{I_o}Q(1/u)L(\overline{r_o}/u), \tag{9}$$

where

$$Q(1/u) = \exp[j\pi R^2/(\lambda u)], \tag{10}$$

$$L(\overline{o}/u) = \exp[j2\pi(o_x x + o_y y)/(\lambda u)], \tag{11}$$

and $C_1$ is a complex constant and there is no linear phase associated with a point object in this case as the optical axis is assumed to match with the centres of all the optical elements. The complex amplitude after the binary axicon and PSA is given as

$$\psi_2 = C_1\sqrt{I_o}L(\overline{r_o}/u)Q(1/u)exp(-j\Phi_{\text{DOE}}), \tag{12}$$

where $\Phi_{\text{DOE}} = \Phi_{\text{Axicon}}$ when an axicon with concentric rings is used and $\Phi_{\text{DOE}} = \Phi_{\text{PSA}}$ when an axicon with sieve configuration is used. The intensity pattern observed at a distance of $v$ from the DOE is given as a convolution of the complex amplitude with the quadratic phase function $Q(1/v)$,

$$I_v(\overline{r_s};\overline{r_o},u) = \left|C_1\sqrt{I_o}L\left(\frac{\overline{r_o}}{u}\right)Q\left(\frac{1}{u}\right)exp(-j\Phi_{\text{DOE}})\otimes Q\left(\frac{1}{v}\right)\right|^2, \tag{13}$$

where, '$\otimes$' is a 2D convolution operator. From Eq (13), if the $\Phi_{\text{DOE}}$ equals $(\pi R^2/\lambda)(1/u + 1/v)$, then in the observation plane a focused point is obtained as the other phase components are cancelled. But for an axicon, the behaviour is quite interesting as it has constant radial spatial frequency. Within the focal depth of the axicon, there is always a radial region of the axicon which has a phase distribution same as $(\pi R^2/\lambda)(1/u + 1/v)$. This region is responsible for generating the central maximum of the Bessel function and the other radial regions which does not match the above phase distribution generate ring patterns around the central maximum. Higher the phase difference, the larger the diameter of the ring and the smaller is the energy density [25]. The intensity for a 2D object $O$ can be expressed as $I_o=O\otimes I_v$. Unlike a coherent source, where the complex amplitude is convolved, here only the intensity distribution is convolved as there is no spatial coherence present to generate the interference terms.

The images of the simulated axicon and PSA with same period are shown in Figs. 2(a) and 2(b) respectively. The images of the diffraction pattern at the same distance for axicon and PSA are shown in Figs. 2(c) and 2(d) respectively and their normalised intensity profiles are shown in Figs. 2(e) and 2(f) respectively. There are interesting observations when comparing the diffraction patterns of axicon and PSA. The central maximum is sharper for axicon than PSA and the sidelobes of axicon appear less strong than that of the PSA. The two intensity distributions were normalised over the maximum of both and the intensity profile is compared as shown in Fig. 2(g). It is seen that axicon has a higher contrast of central maximum to the surrounding rings compared to that of PSA and the full width at half maximum (FWHM) obtained for axicon is smaller than that of PSA. This behaviour is different from the observations made with a Photon sieve Fresnel lens [26], where an improvement in the resolution was observed. The difference may be accounted to the difference between the areas of the two-phase levels. In the previous simulation studies, the areas were made equal to obtain the maximum efficiency in addition to maintaining the phase difference between two levels at $\pi$ for a two-level structure [2]. Due to the replacement of rings by circular discs, this condition cannot be met accurately. A previous study on PSA directly fabricated on the tip of optical fibre exhibited characteristics of an axicon [27]. To further evaluate the focal characteristics, the axial characteristics are studied for axicon and PSA under ideal conditions. The axial variation of the intensity distributions at ($y$=0) line for axicon and PSA are shown in Figs. 3(a) and 3(b) respectively. Comparing the above two figures, it is seen that the performances are similar. From the simulation studies, it is seen that approximating a binary axicon using circular discs does not yield a better performance but reduces the contrast of the central maxima with respect to the surrounding ring patterns. However, the fabrication of PSA using femtosecond ablation is relatively easier and the fabrication of a conventional axicon.

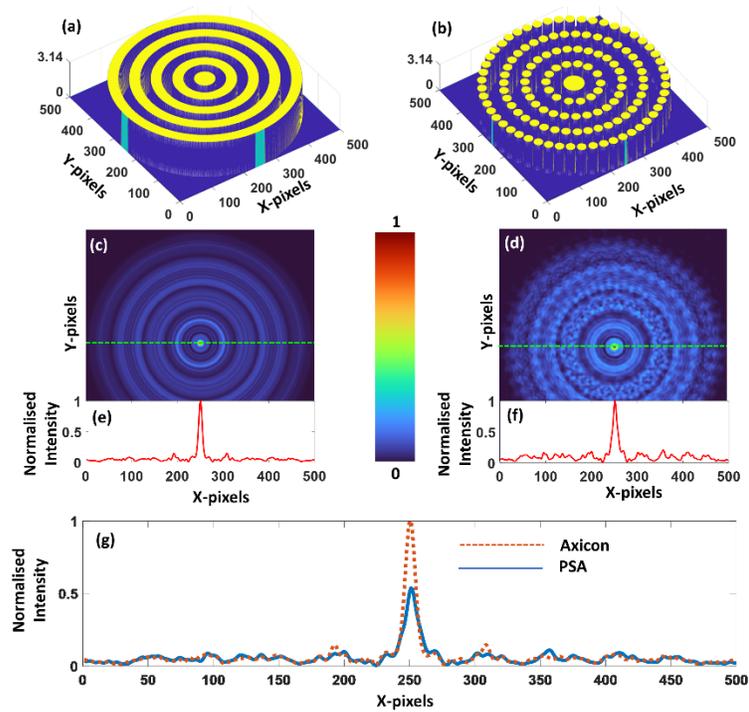

Fig. 2 Phase profile of (a) axicon and (b) PSA for a simulation wavelength $\lambda = 617$ nm. Intensity distribution obtained for (c) axicon and (d) PSA. Locally normalised intensity profile of the line data ($y=0$) for (e) axicon and (f) PSA. (g) Universally normalised intensity profile of the line data ($y=0$) for axicon and PSA.

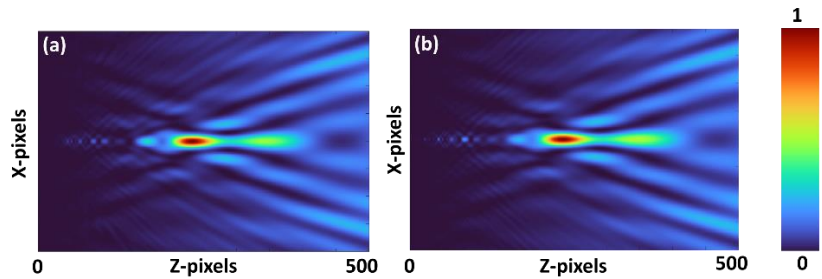

Fig. 3 Locally normalised axial intensity distribution along the line ($y=0$) for (a) axicon and (b) PSA.

The next step is to study the imaging performances of axicon and PSA using a test object. All experiments for this study have been carried out at Nanolab, Swinburne University and so the test object has been selected as the letters of "Nanolab" in arial font as shown in Fig. 4(a). The image of the object obtained using a Fresnel lens, axicon and PSA are shown in Figs. 4(b)-4(d) respectively. It is seen that the sidelobes generated by axicon affects the imaging quality drastically as expected and the results of PSA are distorted more than a conventional axicon. A computational image processing using the non-linear reconstruction (NLR) is used to clean the image of the effects arising from the surrounding ring patterns [23,28-30]. The cleaned image by NLR is given as $I_R = \left|\mathcal{F}^{-1}\left\{|\tilde{I}_v|^\alpha exp[i\,arg(\tilde{I}_v)]|\tilde{I}_o|^\beta exp[-i\,arg(\tilde{I}_o)]\right\}\right|$, where $\alpha$ and $\beta$ are tuned to obtain the lowest entropy. The images generated by axicon and PSA after applying NLR is shown in Figs. 4(e) and 4(f) respectively. To show the difference between the use of coherent and incoherent illumination, the results obtained coherent light source for an axicon and PSA are shown in Figs. 4(g) and 4(h) respectively. Comparing the results obtained for spatially incoherent light with coherent light, the quality of images is better for spatially incoherent illumination than coherent due to the lack of interference.

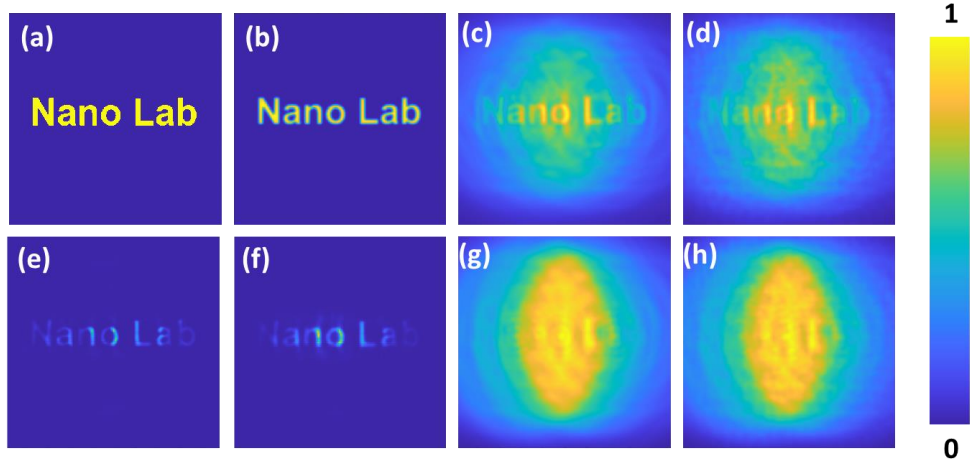

Fig. 4 Image of the (a) test object. Image of the object generated by a (b) Fresnel lens, (c) axicon and (d) PSA. Computationally cleaned image by applying NLR for (e) axicon and (f) PSA. The images obtained for coherent illumination for (g) axicon and (h) NLR.

Another important observation is that the intensity distribution of axicon and PSA are similar in the central region while they are quite different at the outer regions. The random change in the location of the discs and the spacing between the discs at every consecutive zone disturbs the diffraction pattern resulting in the interesting intensity distribution for PSA. To understand the axial characteristics of axicon and PSA, the intensity distribution $I_v$ was simulated for different distances of $v$ from $v$-$\Delta$ to $v$+$\Delta$ and cross-correlated with the intensity distribution at $v$ as $I_v* I_{v\pm\Delta}$, where '*' is a 2D correlation operator. The intensity distributions for some distances and the plot of the axial correlation values are shown for both axicon and PSA in Fig.

5. The case for a diffractive lens is shown for reference. From this study, it seems that a PSA has a higher axial resolution than an axicon. The randomness associated with the distribution of disc suppresses the peaks around the central maximum. From an imaging point of view, the PSA has a better performance than an axicon. When imaging a thick object and focusing a plane of that object, the information from other planes will be of slightly lower intensity in the case of PSA when compared to an axicon.

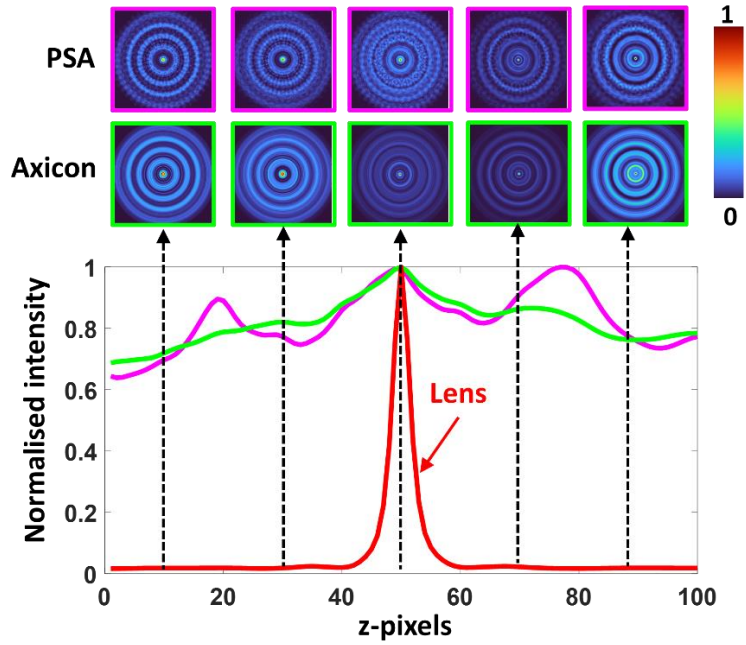

Fig. 5 Plot of $I_v * I_{v\pm\Delta}$ for different distances for axicon (pink), PSA (green) and a lens (red). The intensity distributions for axicon, and PSA for some distances have been provided.

## 3. Experiments

### 3.1 Fabrication

The fabrication was carried out on sapphire substrate with a thickness of 500 μm and index of refraction $n_s = 1.76$. In order to have π phase difference between the two levels the thickness needed to be milled is given as $t = \lambda/2(n_s-1)$ which is ~0.41 μm for $\lambda = 617$ nm which is the wavelength of the source used for optical testing. The substrate was cleaned prior to fabrication with acetone and iso-propyl alcohol (IPA) and blown dry. Three optical devices, namely an axicon, a PSA and a sparse PSA were fabricated. The fabrication was carried out using a Pharos laser (Light Conversion, Lithuania) operating at 200 kHz repetition rate, $\lambda = 1030$ nm wavelength, 2.5W average power, 230 fs pulse duration and 5× magnification, NA = 0.14 numerical aperture Mitutoyo Plan Apo NIR infinity corrected objective. Two pulse bursts per ablation spot was used (with an exception for the axicon that did not use a pulse burst) in combination with pulse energy after the objective attenuated to 5 µJ to achieve the design ablation depth and the ablated spot diameter of d = 8.9 µm. The diameter of ablated spot was closely matching the diameter for the used NA: $d = 2w = 1.22\lambda/NA$, where $w$ is the waist of the beam at focus. The peak intensity of a 5 µJ pulse (peak intensity is twice the average) is 69 TW/cm$^2$ at the fluence of 16 J/cm$^2$. This is markedly larger than the ablation threshold of sapphire at ~2 J/cm$^2$ and ~10 TW/cm$^2$. The speed of Aerotech positioning stages was set to 2 mm/s. Stage travel with position sensitive output (PSO) mode was used to ensure accurate pulse spacing. The fabrication code was written in SCA software (Workshop of Photonics Altechna, Lithuania). The code comprised of concentric circle commands with circle radius increasing by 16 µm till the overall design diameter reached 5 mm. Laser pulse spacing in a given circle was set to be 5 µm, 8 µm, and 12 µm for respectively the axicon, the PSA, and the sparse PSA fabrications. The Pharos system unlike electron beam lithography system always works in fixed beam movable stage mode (FBMS) mode which is a special case in lithography systems such as RAITH150[2]. Conventional raster and serpentine scanning approaches are time consuming as every pixel of the design must be scanned. In this study, time efficient circular scanning procedure was selected where the pixel locations were selected for the beam to traverse and the stage was moved only to those locations as shown in Fig. 6(a). The scanning procedure started from the centre and moved gradually outward following the coordinate map. In this way, the fabrication time was reduced by at least half of that of conventional scanning approaches. The fabrication time for the sparse PSA, PSA and axicon was approximately 10 minutes. When the stage moved to the coordinates, a laser pulse (single pass) ablates the corresponding voxels.

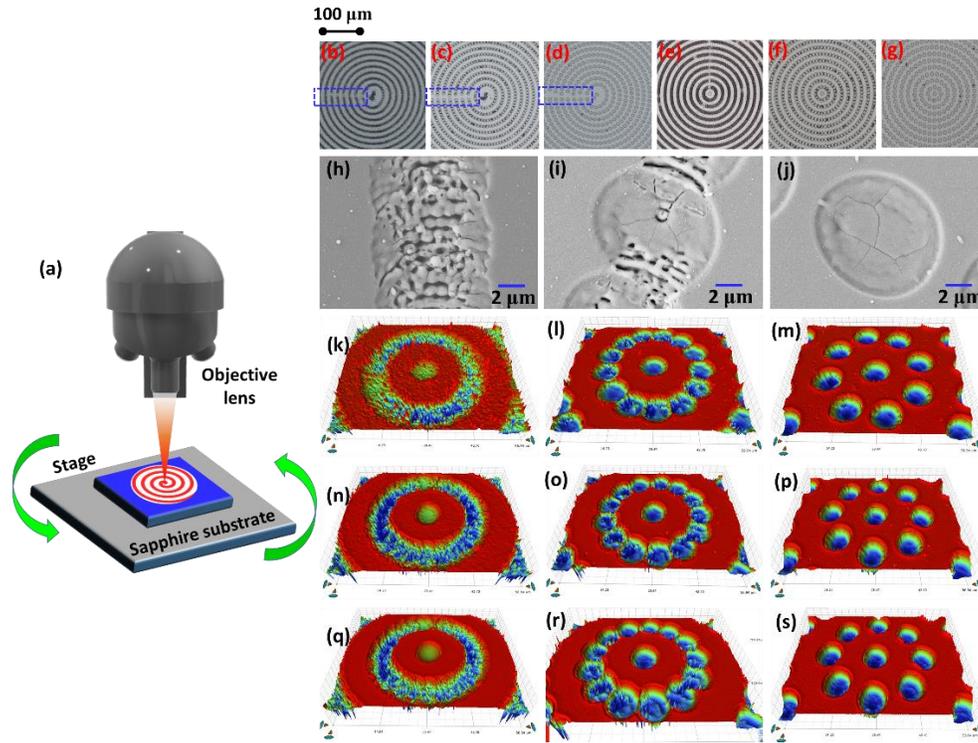

Fig. 6 (a) Schematic of the scanning procedure. Optical microscope images of the central part of (b) axicon, (c) PSA and (d) sparse PSA with stitching and stigmatism errors. Optical microscope images of the central part of (e) axicon, (f) PSA and (g) sparse PSA after the correction of stitching and stigmatism errors. SEM images of a single pulse fabrication for (h) axicon, (i) PSA and (j) sparse PSA; ripples with period ~1 μm ≈ λ are recognizable in the strongly overlapped pulse exposure regions. Surface profiles of (k) axicon, (l) PSA and (m) sparse PSA, after femtosecond fabrication. Surface profiles of (n) axicon, (o) PSA and (p) sparse PSA after ultrasonic cleaning in IPA and surface profiles of (q) axicon, (r) PSA and (s) sparse PSA after KOH etching.

In the first step, the effect of stitching error arising due to the condition that the circumference of the zones is not equal to the integral multiples of $\Lambda/2$. The optical microscope images of the fabricated axicon, PSA and sparse PSA are shown in Figs. 6(b)-6(d) respectively. The stitching error is clearly visible. In addition to the stitching error there was error due to stigmatism in the stage resulting in oval shaped zones. The error compensation terms were introduced in the code to reduce both the stitching error and stigmatism. The optical microscope images after correction are shown in Figs. 6(e)-6(g) respectively. The SEM images of a single disc for axicon, PSA and sparse PSA are shown in Figs. 6(h)-6(j). It can be seen that with the overlap of pulses in PSA and axicon, the surface is damaged more and becomes rough due to ripple formation [31] while the sparse PSA without disc overlap has a smoother and nearly flat surface with only cracks visible at the molten region on the bottom. A fast thermal quenching of shallow melt pool is the most probable cause of formation of nano-cracks; note the melting temperature of sapphire is 2040°C. A remarkable feature of the used two-pulse ablation was a flat bottom of the ablation pit. The apparent melting was contributing to the flatness of ablation site; however, the other contribution should be in the profile of energy deposition. The die-met transition, when dielectric is transferred to the metallic plasma state with permittivity decreasing to $\varepsilon = n^2 = 0$ at the dielectric breakdown [32], is defining the lateral absorption profile. For a Gaussian-like intensity at the focus, the most intense central region is turning into the most reflective during the laser pulse, hence, less energy is deposited to the sample. This qualitatively explains the ablation profile which would be expected for a hat-top intensity profile (rather than the Gaussian). The detailed mechanism of ablation is out of scope of this study focused on optical performance, however, is worth of a deeper investigation.

The rough surfaces of PSA and axicon are expected to scatter light more resulting in lower efficiency in the first diffraction order. After fabrication, it was noted that there was redeposition of debris both on the milled and unmilled areas of the device. Two step cleaning procedures were carried out. After femtosecond fabrication, the sample was cleaned in IPA in the ultrasonic bath for 5 minutes. The surface profiles of the axicon, PSA and sparse PSA before and after cleaning are shown in Figs. 6(k)-6(p) respectively. After the first cleaning, most of the debris are removed. The next cleaning was carried out with a KOH solution with (2M concentration for 2 hrs) to remove the redeposited amorphous Sapphire and there was further improvement rendering an even smoother profile [33] for axicon, PSA and sparse PSA as shown in Figs. 6(q)-6(s). The average depth values measured for the axicon, PSA and sparse PSA were 360 nm, 416 nm, 420 nm respectively. This required depth for a π-phase shift was achieved in a two-pulse irradiation without change of the focal position. The Rayleigh length (half of depth-of-focus DOF) is $z_R = \pi w^2/\lambda = 61.4$ μm. This large DOF increase tolerance of high-fidelity fabrication over large areas when some tilt of a surface plane can be expected. The dark regions in the microscope images and the higher roughness in surface profiles indicates that there was some amount of redeposition of the material into the ablated area due to the overlap of the voxels (volume element, i.e., a 3D pixel) during fabrication. This redeposition is dependent upon various factors such as the pulse energy, material characteristics such as melting point, crystalline or amorphous.

*3.2 Optical testing*

The optical testing was carried out using a high-power LED from Thorlabs (M617L3, $\lambda_c$ = 617 nm, full width at half maximum (FWHM) = 18 nm) and a spectral filter was used at 600 nm with a width of 10 nm to improve the temporal coherence. A pinhole with a size of 100 μm and a cross-shaped object were used for imaging. A 3× magnifying system was used to reimage the intensity distribution that is close to the DOEs on an image sensor (Thorlabs DCU223M, 1024 pixels × 768 pixels, pixel size = 4.65 μm). The DOEs were mounted one after the other and the intensity distributions were recorded. The intensity distributions recorded at a distance of 5 mm for the axicon, PSA and sparse PSA are shown in Figs. 7(a)-7(c) respectively. The cross object was then mounted instead of the pinhole and the intensity patterns were recorded at 3 cm from the DOEs. The images of the cross object recorded for the axicon, PSA and sparse PSA are shown in Figs. 7(d), 7(f) and 7(h) respectively. The images were cleaned using the images obtained for a pinhole and the NLR method and the cleaned images are shown in Figs. 7(e), 7(g) and 7(i) respectively. To improve the cleaning results, additional filters such as median filter and correlation filters were used as demonstrated in [34]. The images after cleaning for axicon, PSA and sparse PSA appear similar.

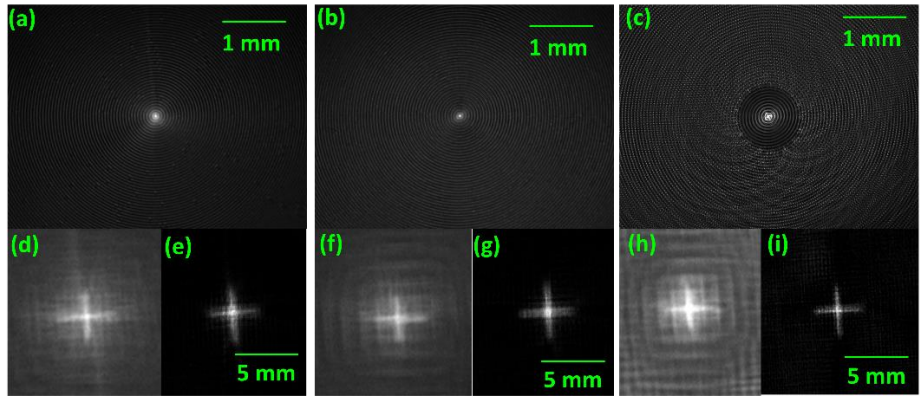

Fig. 7 Intensity distributions recorded for a pinhole at distances 5 mm for (a) axicon, (b) PSA and (c) sparse PSA. Recorded intensity distributions for a cross shaped transmission object at 5 mm for (d) axicon, (f) PSA and (h) sparse PSA. Cleaned images of cross for (e) axicon, (g) PSA and (i) sparse PSA.

Therefore, it is feasible to use a sparse axicon instead of a regular axicon for imaging applications. By virtue of a wide spectral transparency of sapphire from UV to IR, optical elements with ~1 cm cross section can be made within ~10 min which allows for a wider use of such prototyping for more complex optical systems. It must be noted that the cleaning procedure does not require the recording of the intensity distribution for a pinhole exactly at the same location as the object owing to the high focal depth of the imager. For the same reason, while cleaning images of thick objects the planes that are out of focus will have a higher intensity as the plane in focus. Consequently, if the information of two planes overlap laterally then the information could not be perceived accurately. A recording at a plane can be used to clean the image for many distances. The image cleaning results for the intensity distributions recorded for the cross object at 5 cm (Fig. 8(a)) and 6 cm (Fig. 8(b)) using pinhole intensity distributions recorded at 5 mm are shown for axicon and PSA as shown in Fig. 8(c) and 8(d) respectively. Since the cross object is a simple object, the cleaning may not appear significant. A synthetic object consisting of the letters 'APHB' was used next and the corresponding intensity distributions for PSA were synthesized by convolving the intensity distributions recorded for a pinhole with the object function. The intensity distribution of the synthetic object at two planes 5 cm and 6 cm are shown in Fig. 8(e) and 8(f) respectively. The cleaned images using the pinhole recording at 5 mm are shown in Figs. 8(g) and 8(h) respectively.

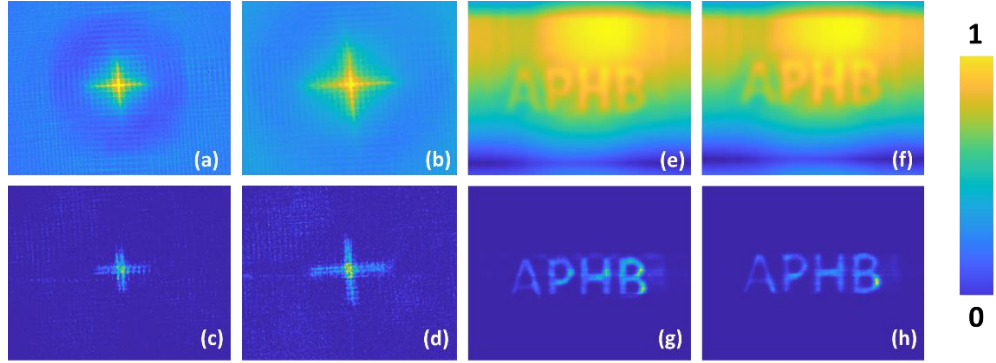

Figure. 8 Intensity distribution of cross object recorded at (a) $z = 5$ cm and (c) $z = 6$ cm and the corresponding cleaned images (c) $z = 5$ cm and (d) $z = 6$ cm respectively. Intensity distribution of synthetic object 'APHB' recorded at (e) $z = 5$ cm and (f) $z = 6$ cm and the corresponding cleaned images (g) $z = 5$ cm and (h) $z = 6$ cm respectively.

4. **Conclusion**

Rapid fabrication of large area DOEs using the femtosecond laser fabrication system has been investigated. A simpler design with the function of a binary axicon was used for the study. The fabrication time was only 10 minutes using the femtosecond fabrication method for a large area of 5 mm × 5 mm. The binary axicon was realized in three configurations namely conventional axicon, PSA and sparse PSA. It was noticed that when the beam overlaps during milling results in redeposition of material resulting in a lower depth than the case without beam overlap which goes against the common belief that overall higher exposure with beam overlap increases the depth. In addition, the case with beam overlap has a higher roughness value than the case without beam overlap. This increase in roughness is partly contributed by the redeposition and partly due to light-matter interaction at temperature changes caused by ablation by the previous pulse. Reduction of debris and mechanism of ripple-free deep ablation at high irradiance which exceeds ablation threshold more than an order of magnitude are directions for future investigations which were not practical at low repetition rates [35].

One interesting outcome of this study is that it seems it may be necessary to approximate the design of the DOE functions as in this case where a sieve configuration was used instead of rings to achieve a milling favorable design. As is seen in this study, the intensity distributions obtained for the sieve configuration is similar to that of the conventional axicon with a slight loss of lateral resolution and increase in axial resolution. Therefore, it should be possible to modify the design of DOEs without compromising the diffraction performances and at the same time achieve high fabrication accuracy without material redeposition and increase in surface roughness.

Some of the recent studies on controlling the distribution of debris using externally applied electric and magnetic fields offer opportunities to control the light-matter interactions and achieve better fabrication results [36-37]. We believe that the direction of research will enable the fabrication of large area structures suitable for incoherent illumination and astronomical imaging applications. With the introduction of spatial multiplexing [38], optimization [39] and redesigning of DOEs, the performance of the femtosecond fabrication system can be improved further. Astronomical imaging is one of the areas which require large area optical devices with capabilities to perform a high signal to noise ratio. We believe that the femtosecond ablation based rapid fabrication and computational optics in tandem will lead to the development of advanced astronomical imaging technologies. Some of the latest developments in astronomical spectral imaging technologies such as FOBOS [40] require numerous micro-optical devices for the successful implementation of free space to fiber bundle coupling for spectral imaging. We believe that the current work consisting of rapid fabrication and beam cleaning will support retrieving spatial information in addition to the recorded spectral information [41].

**Funding.** ARC LP190100505 is acknowledged for funding.

**Acknowledgments.** We are grateful to the Workshop of Photonics, Ltd. for femtosecond laser fabrication system acquired via technology transfer project. We acknowledge funding for the Nanolab by Swinburne University of Technology (SUT). DS is grateful for support via Honors and MH via PhD programs at SUT.

**Disclosures.** The authors declare no conflicts of interest.

**Data availability.** The datasets used and/or analysed during the current study are available from the corresponding author on reasonable request.

**List of abbreviations:** PSA – Photon sieve axicon; DOE – Diffractive optical elements; FWHM – Full width at half maximum; NLR – Non-linear reconstruction; IPA – Isopropyl alcohol; LED – Light emitting diode; DOF – Depth of focus; M – Molar; KOH – Potassium hydroxide; SEM – Scanning electron microscope; UV – Ultraviolet; IR – Infrared; FBMS – Fixed beam movable stage; M – Molar; NIR – Near infrared; NA – Numerical aperture; PSO – Position sensitive output.

**Author contributions:** Light matter interaction and dynamics conceptualization – S. J; Fabrication – D.S, S.H.N, M. H and T. K; Characterisation – M. H, D. S, S.H.N, V. A and T. K. KOH etching – S. H. N, V. A and D. S; Simulation – V. A and D. S; Optical testing – V. A, S. H. N and D. S; Manuscript writing – V. A and S. J; Review, editing and proof reading – all the authors; Project guidance – K. G, S. J; Resources – S. J; Funding – S. J.